\documentclass[12pt]{article}
\usepackage{amssymb}
\usepackage{amsmath}

\def\beq{\begin{equation}}\def\eeq{\end{equation}}
\def\bea{\begin{eqnarray}}\def\eea{\end{eqnarray}}

\begin{document}
 
\title{Can Bohmian particle be a source of "continuous collapse" in GRW-type theories}
 
\author{Roman Sverdlov
\\Raman Research Institute,
\\C.V. Raman Avenue, Sadashivanagar, Bangalore -- 560080, India}
\date{March 15, 2011}
\maketitle
 
\begin{abstract}

\noindent The purpose of this paper is to unite Pilot Wave model with GRW ideas through a proposal that Bohmian particle serves as a source of continuous collapse. The continuous trajectory of a particle allows the particle-centered collapse mechanism to be continuous as well. This allows us to remove the "stochastic" element from typical GRW proposals. 

\end{abstract}
 
\subsection*{1. Introduction}

According to Bohmian interpretation of quantum mechanics (see \cite{Bohm1} and \cite{Bohm2}), the wave and the particle co-exist as two separate substances. The wave evolves according to Schrodinger's equation, while particle moves according to \emph{guidance equation} given by
\beq \frac{d \vec{x}}{dt} = \frac{1}{m} \; \vec{\nabla} \; Im \; ln \; \psi \eeq
Suppose at the beginning of the universe, a "biased coin" was tossed in order to determine where to place a particle in space. If the bias of a coin happened to be $\rho = \vert \psi \vert^2$, then the "probability current" ($\vec{J}$) associated with \emph{classical} probability ($\rho$) of finding a particle will coincide with the "probability current" ($\vec{j}$) associated with $\vert \psi \vert^2$ that we obtain from Schrodinger's equation. For this reason $\rho$ and $\vert \psi \vert^2$ will continue to coincide at the time $t+ dt$. Then, for the same reason, they will coincide at $t+2dt$, and so forth. This is the key argument in favor of Born's rule arising out of Bohmian model. 

According to the Bohmian theory, the collapse of wave function is a result of wave function splitting into non-overlapping branches. In such scenario, a particle will "happen" to occupy only \emph{one} of them, and the zero-probability regions will prevent it from going back to any other branches. These branches are interpreted as different outcomes of a measurement. In this language, the "collapse" of the wave function is equivalent to these branches disappearing. According to the theory, they do not disappear and, therefore, no true collapse occurs. But, due to their lack of overlap, they have no effect on a Bohmian particle. Thus, the particle "thinks" they disappeared, which results in \emph{appearance} of collapse of wave function. This phenomenon is commonly referred to as "effective collapse".

However, the reliance on the claim of "lack of overlap" raises several problems. First of all, the fact that branches do not overlap was disputed by Leggett. According to his argument, the only thing we know is the emergence of sum rule, but the latter implies the \emph{lack} of evidence of the \emph{absence} of a "split" which is \emph{not} the same as the evidence of its presence (see \cite{Leggett1} and \cite{Leggett2}). Secondly, even if we believe that the splitting of the branches occurs, the particle can "cross" that "gap" once the theory is non-local in configuration space.

 Such "non-local" theories have been recently proposed in \cite{Epstein} and \cite{Sverdlov}. Both \cite{Epstein} (Chapter 3) and \cite{Sverdlov}, in fact, acknowledge the "traveling between universes" that would occur as a result of this non-locality (although in \cite{Sverdlov} rather artificial mechanism was presented in "getting rid" of these unwanted universes).   Finally, if one hopes to introduce gravity into the the theory (which is beyond the scope of this paper) the non-linear nature of gravity might imply non-zero gravitational interaction between these branches.  

As Leggett pointed out (see \cite{Leggett1} and \cite{Leggett2}) the above problems are completely absent in GRW theory. The GRW theory postulates "real" collapse which is is \emph{exclusively} based on the distance in configuration space (and, therefore, no appeal to the shape of the curve, including the "branching" is made). The "real" collapse also would get rid of any unwanted branches \emph{if} such were created (although, of course, if they are not created to begin with, this is "even better"), so the issues such as "particle crossing the gap between the branches" doesn't arise either. 

The jist of the theory is that a wave function is being periodically multiplied by Gaussians around random points at random times; these events are called "hits". These Gaussians are very wide, and occur very rarely. Thus, their effects on few particle systems are negligible. In multi-particle case, however, the entanglement implies that any particle is being affected by "hit" performed on any other particle. This makes the number of "hits" it experiences far more frequent, which makes the degree of localization significant. The same is true for external particle having short term interaction with the multiparticle system, such as electron interacting with the screen.

However, GRW theory is stochastic in its original form, which is in conflict with the goal of some physicists (including the author of this paper) to come up with deterministic theory. There have been "continuous spontaneous localization models" (see \cite{continuous}) that postulated extra classical fields that "trigger" the above described localization behavior. Thus, instead of random "hits", the "multiplication by Gaussian" is continuous and its (very slow) rate varies according to the behavior of the above mentioned classical field. This enables the theory to be deterministic. 

In this paper, we will propose a different alternative of imposing determinism upon GRW model. Instead of claiming that the "classical trigger" involves these fields, I will claim that the Bohmian particle, itself, triggers a "continuous localization" around itself, as a center. In other words, the evolution of wave function is given by 
\beq \frac{\partial \psi}{\partial t} = i H \psi - a (\vec{x} - \vec{x}_B)^2 \eeq
where $\vec{x}_B$ is a location of "beable particle". Upon integration over finite time interval, the above equation also produces the multiplication by Gaussian. The key difference between this dynamics and GRW model is that in our case we no longer have discrete hits, which were the only sources of randomness. 

There are different ways of summarizing what we are proposing to do. On the one hand, we can say that we plan to use GRW as a "supplement" to Bohmian dynamics that "gets rid" of unwanted parts of wave function (and, therefore, renders the argument about "splitting into branches" unnecessary). At the same time, it can also be viewed as Bohm supplementing GRW in a sense that the latter no longer needs to postulate extra classical fields as "sources" of localization. In the former case we claim that GRW is "a little addition" to the "main Bohmian context" while in the latter case we claim just the opposite. We can also think of it in "unbiased" way and say we are simply "merging" Bohm and GRW into a single "hybrid" between the two. This is the preferred view of the author. 

However, the author of this paper happens to have some other, unrelated, objections against GRW. The first objection is that the "multiplication by Gaussian" goes against momentum conservation. The second, and unrelated, subjection is that the "smooth dynamics" summarized in the above equation can be re-phrased in terms of "imaginary Hamiltonian" $H = -iax^2$. While neither of these issues affect the validity of the main point of the paper, we will still attempt to address those in Chapters 6 and 7, respectively, for the sake of completeness of presentation. 

\subsection*{2. The problems created by particle's influence on the wave}

One key implication of coupling Bohm to GRW is that the wave-particle interaction happens \emph{both} ways. This should be contrasted with purely Bohmian case when the wave influences a particle and the particle has no effect on the wave. The \emph{lack} of the influence of a particle on the wave is precisely the reason why unwanted branches continue to exist. In our case, the \emph{presence} of that influence is claimed to get rid of these branches.

However, once we allow a particle to influence a wave, one serious question arises. In order to write down Born's rule, we have to make two key assumptions. One is that we \emph{do} know the exact value of $\psi$ and the other is that we do \emph{not} know the exact location of a particle. If we did not know the value of $\vert \psi \rangle$, we would not be able to use it in Born's rule. On the other hand, if we knew the location of a particle, the "probability distribution" would have been a $\delta$-function, in violation of Born's rule. This raises a question: how can we "know" $\psi$ despite the fact that it is being influenced by $\vec{x}$ which we do \emph{not} know? 

This question can also be asked in another way. Suppose we don't know $\psi$ at any other time, but we are "told" what $\psi$ is at time $t=t_0$. We are now trying to estimate the probability of finding a particle at different locations in $\vec{x}$ at this particular time. Now, \emph{if} a particle happens to be located at $\vec{x} = \vec{x}_0$, then we can do "reverse dynamics" and say that, at time $t- \delta t$, the wave function was $\psi +iH \psi \delta t - G(\vec{x}_0) \psi$, where $G(\vec{x}_0)$ is an "additional influence" the particle exerts onto a wave function. 

Now, if we again look at the "forward dynamics", in order for particle to "end up" being at $\vec{x}_0$ we have to \emph{first} produce $\psi +iH \psi \delta t - G(\vec{x}_0) \psi$ and \emph{if} a particle \emph{happens to be} at $\vec{x}_0$, then this would ultimately produce the situation we are looking for after the interval $\delta t$ passes. This means that if it happens that, for the time $t_0 - \delta t$, the state $\psi +iH \psi \delta t - G(\vec{x}_1) \psi$ is "more likely" to be produced than $\psi +iH \psi \delta t - G(\vec{x}_2) \psi$, then at time $t_0$, the particle is "more likely" to be found at $\vec{x}_1$ than at $\vec{x}_2$ at the "ideal" situation that $\vert \psi (\vec{x}_1) \vert ^2 = \vert \psi (\vec{x}_2)^2$.

Of course, it is \emph{possible} that the probability of producing $\psi +iH \psi \delta t - G(\vec{x}_0) \psi$ at time $t_0 - \delta t$ is independent of the choice of $\vec{x}_0$, but we do not \emph{know} that. Finding out whether or not that is the case is quite difficult since the probability of production of $\psi +iH \psi \delta t - G(\vec{x}_0) \psi$ depends on the situation \emph{before} $t_0 - \delta t$ which means that we can not use the "back dynamics" from time $t_0$ to assist ourselves. On the other hand, the "forward" dynamics for the times $t < t_0 - \delta t$ depends on the entire quantum mechanics (or, in non-relativistic case, quantum field theory) to which $\psi$ is subject to, which makes the proof of any kind of assertion extremely difficult. 

One way to illustrate the difficulty is to consider a problem of a kid and a candy. We are told about the amount of candies in different locations and we are asked to find out probabilities at which a kid is to be found at each of these places. On the one hand, one can argue that kid is more likely to be found where there is more candy, since he is "drawn" to the candy. On the other hand, we can argue that he is more likely to be found where there is \emph{less} candy since he would eat the candy (and, thereby, decrease its number) at whatever location he occupies. 

In order to answer this question we need to have an \emph{additional knowledge} about the way candy is distributed. If we happened to know that for some reasons \emph{independent of the child} candies are typically distributed evenly, we would conclude that a kid is to be found where there is \emph{less} candy. After all, the only conceivable reason there is "less candy" to begin with is that he ate them. On the other hand, if we are told that, for some reason, the distribution of candies is \emph{highly uneven} then the kid would be more likely to be found where there is \emph{more} candy. After all, the "unevenness" of candy distribution is so high that a kid can't possibly eat enough candies to "undo" it. Thus, he would arrive at the place with "more candies", eat some, and there would \emph{still} be "more candies than average" after he is done eating. 

Furthermore, even if we do know that candies are distributed evenly, the situation might not be as simple. For example, if the "even distribution of candies" happens very rarely, then the kid would eat candies from all the rooms, resulting in their uneven distribution; afterwards, he would "move" to the room where more candies are "left". On the other hand, if the candies are being "evened out" by outside sources quite often, he won't have time to eat candies from all the rooms (provided that we assume that there are a lot of rooms) and, therefore, the room he would occupy would be one of the "few" rooms he "had time" to reach and, therefore, would have "fewer candies". But, at the same time, he might have still had time to reach more than one room and, between the rooms that he could reach, he could "pick" the one where he ate less candies since more candies are "left" for him to eat.  From the point of view of this picture, the probability of finding a kid in \emph{either} a room with a lot of candies \emph{or} a room with very few ones is very small, but for very different reasons. The probability "peaks" around the rooms with "fewer but not too few" candies, making the theory "non-linear". 

We should notice that the non-linearity of the above paragraphs was produced within "black or white" context of "even distribution of candies". The situation becomes even more complicated if we allow the distribution of candies to be uneven, but, at the same time, not uneven enough to "qualify" for the "other extreme" discussed earlier. Thus, we could not blindly say that the room with more candies attracts a kid since it is \emph{possible} he eats more candies than the "standard deviation" of their distribution; but, at the same time, such criteria might "partly" hold since standard deviation is quite large and a kid \emph{might well} eat fewer candies than that.  

Going back to physics, the "kid" is analogous to a "particle", a "candy" is analogous to the wave, "kid wanting to go where there is more candy" is analogous to "wave guiding the particle through Pilot Wave model" and "kid eating the candy" is analogous to the particle influencing the behavior of a wave through GRW-type phenomenon centered around its own location. The "way in which candies are distributed" is analogous to the knowledge of quantum field theory. Thus, the fact that we need to know the way candies are distributed in order to decide how to estimate the probabilities of finding a kid implies that we need to know quantum field theory in order to estimate the probability of finding the particle. This is in sharp context with standard quantum mechanics (or field theory) where, in principle, we can estimate the probability of finding a particle simply by "being told" about the probability amplitude, even if we, ourselves, do not know quantum mechanics and can not reproduce it. 

Apart from mathematical difficulties involved, there is also a conceptual one. If we pretend for a moment that a particle doesn't exist and, therefore, wave evolves without its influence, it might still be "reasonable" to say that some shapes of waves are "more likely" to be produced than other shapes. But, at the same time, the unitary evolution is completely deterministic. Thus, in order to talk about different probabilities of various outcomes, we need to be uncertain about initial conditions. Technically, we could describe the uncertainty of wave function through density matrix. But additional notation is not going to answer our conceptual question: just what "makes" one kind of wave function "more likely" than the other kind? These questions is one of the main things I am attempting to tackle in this paper. 

\subsection*{3. The "bigger" space consisting of objects of the form $(\vert \psi \rangle, \vec{x}$) }

I propose to address the above questions in the following way. In terms of kid and a candy I would combine them into a single object, $({\rm kid}, {\rm candy})$ that undergoes certain dynamics. Here, by "candy" we mean the \emph{set} of numbers that list all the rooms and list the amounts of candies in every single one of them, and by "kid" I mean a single "number" that tells which room a kid occupies. Then, I would be able to \emph{first} find probability density distribution around any given point $({\rm kid}, {\rm candy})$ and then, later, find a \emph{conditional probability} of kid being in a certain place for a specific distribution of candies. If "kid" is viewed as "$x$-coordinate" and "candy" as "$y$-coordinate", it makes perfect sense to ask about the conditional probability of finding $x$ (that is, a kid) for afore-specified $y$ (that is, candy distribution). This will allow us to use "candy" as a way of determining a "probability" of finding a kid, just like $\psi$ determines a probability of finding a particle. 

By the similar logic, I will combine a state $\vert \psi \rangle$ and a particle configuration $\vec{x}$ into a single object, $(\vert \psi \rangle, \vec{x})$ which undergoes random walk in that "bigger space". Thus, $(\vert \psi \rangle, \vec{x})$ will be a \emph{single point} as far as the "bigger space" is concerned. That "bigger space", however, will have $\infty^{3N} + 3N$ dimensions. In particular, we need $3$ dimensions in order to describe one point particle, and, therefore, $3N$ dimensions to describe configuration of $N$ of them. Furthermore, we need $\infty$ dimensions to describe wave function over one dimensional space and, therefore, $\infty^{3N}$ dimensions in order to describe wave function over $3N$ dimensions. But if we combine all of these into a single $\infty^{3N} + 3N$-dimensional space, we would still be able to think of $(\vert \psi \rangle, \vec{x})$ as a single "point" undergoing random walk in that space. 

In the standard situation where only $\vec{x}$ was undergoing "random walk" there was an "outside influence" by $\psi$. Thus, there was a \emph{dynamic equilibrium} $\rho = \vert \psi \vert^2$, and both sides were time dependent. In our new situation, however, there is nothing "outside" $(\vert \psi \rangle, \vec{x})$. Thus, the equilibrium is static one. In other words, we \emph{first} come up with evolution equation, 
\beq \frac{d (\vert \psi \rangle, \vec{x})}{dt} = u (\vert \psi \rangle, \vec{x}) \eeq
and then determine the time derivative of "classical" probability $\rho (\vert \psi \rangle, \vec{x})$ based on the above dynamics. Finally, we determine what $\rho$ is based on the demand that its time derivative is zero. After that, for any given $\vert \psi \rangle$ I will determine a \emph{conditional probability} of finding $\vec{x}$ at any given point based on 
\beq \frac{\rho (\vert \psi \rangle, \vec{x})}{\int d^{3N} y \; \rho (\vert \psi \rangle, \vec{y})} \eeq
where we have taken the $\vert \psi \rangle = \vert \psi_0 \rangle$ "section" of our space, which answers the question of "how do we know $\psi$". The burden of the theory is to prove that the above expression coincides with $\vert \langle x \vert \psi \rangle \vert^2$. This is now a mathematically well defined statement, which one can reasonably expect to answer either affirmatively or otherwise. 

Now, in order to obtain the dynamics out of the above "static" equilibrium, one has to realize that the above probabilities are strictly due to our "lack of knowledge". Thus, in principle we can "find out" some information about the time $t=t_0$ which would "allow us" to predict what would happen at some future time $t>t_0$. In particular, I claim that, if we have we have "found out" that, at the time $t=t_0$, the configuration of particles is $\vec{x}_0$, we  would be to find out a \emph{new} probability density, $\sigma (\vert \psi \rangle, \vec{x} ; t)$ for some "future time" $t$ based on this information. While $\sigma$ is time-dependent, we would have to use time-independent $\rho$ in our calculation, which is the "static" aspect of the theory. 

Let us ask ourselves what is the probability that $(\vert \psi \rangle, \vec{x})$ will happen to reside within the set $S$ at a time $t > t_0$ and then we can "shrink" $S$ around the point we are interested in. Let us denote the evolution under $u$ as $e^{u(t-t_0)}$. Furthermore, let us  define $a ((\vert \psi \rangle, \vec{x} \rangle), S)$ to be "yes or no" function that is equal to 1 if $(\vert \psi \rangle, \vec{x})$ is an element of $S$ and zero otherwise. Then the probability of finding $(\vert \psi \rangle, \vec{x})$ inside $S$ at time $t$ is given by
\beq \int_{S} [{\cal D} \psi ] d^{3N} x \; \sigma (\vert \psi \rangle, \vec{x}, t) = \int [{\cal D} \psi ] \rho (\vert \psi \rangle, \vec{x} = \vec{x}_0) a (e^{u(t-t_0)} (\vert \psi \rangle, \vec{x} = \vec{x}_0); S) \eeq
Likewise, if we, instead, "find out" the value of $\vert \psi \rangle$ at a time $t_0$ \emph{as opposed to} the configuration of particles, we can obtain the probability distribution at time $t >t_0$ through  
 \beq \int_{S} [{\cal D} \psi ] d^{3N} x \; \sigma (\vert \psi \rangle, \vec{x}, t) = \int d^{3N} x \;  \rho (\vert \psi \rangle)= \vert \psi_0 \rangle, \vec{x}) a (e^{u(t-t_0)} (\vert \psi \rangle )) = \vert \psi_0 \rangle , \vec{x}); S) \eeq
The choice "which" we "find out" is based on the philosophy of a reader. If a reader believes that we "see" point particles and "use" that information to "infer" the wave function, he would use the former equation. If he believes we "see" the wave function while point particles are the things that guide its evolution which we don't directly see, he will use the latter equation. Our theory works equally well in both cases. 

\subsection*{4. Repeated observations}

As we have seen in the previous chapter, the probability $\rho$ is static, while the probability $\sigma$ is dynamic. Its time-dependence depends on our ability to "find out" an additional information at any given time. We have, however, limited the "additional information" we are allowed to find out. In particular we considered an option of "looking at $\vec{x}$" without being allowed to "look at $\vert \psi \rangle$" or visa versa. While, as we said, the choice between these two options is not important, the \emph{existence} of well defined limitation is. After all, if we could measure the exact value of both $\vert \psi \rangle$ and $\vec{x}$ then the determinism would imply the exact prediction for future; in other words, $\sigma$ would be $\delta$-function, which is not what we want.

Now, if we \emph{do} have some limitations on what we are allowed to observe, What \emph{is} important, however, is that we have to be "consistent" in the kind and amount of information we are allowed to "find out". After all, that is what the "consistency" of $\sigma$ depends upon. This raises an important question: does the performance of repeated observations allow us to "infer" extra information and thus ruin this consistency? The answer to this question is that the only way to access past observations is to "look at" our \emph{current} memory. Since our brain is a physical system, our current memory is recorded both in \emph{current} $\vert \psi \rangle$ and $\vec{x}$. Which still reduces to only one observation. 

Now, the idea that in the past we never observed anything and just have a "false memory" that we did is, of course, very unpleasant. A way to go around it is as follows. Yes, we \emph{do} make several different observations over time. But every time we make an observation we "forget" what we have seen once a small time interval $\delta t$ passes by. Thus, at a time $t_1 + \delta t$ we no longer remember anything we have seen at $t_1$. Then, at time $t=t_2$ we make another observation. This observation included the configuration of particles in our brain. From the configuration of particles in the brain we can \emph{infer} some aspects of what we have observed at $t=t_1$. But, at the same time, we \emph{still} don't remember that observation; we are simply \emph{inferring it} from observing our brain at $t= t_2$. 

It is interesting to note that if we \emph{did} have direct access to our observation at $t=t_1$ then we would indeed get additional clues about the system that we can not obtain from the single observation. At the same time, by observing our brain at $t=t_2$ we can \emph{not} get these clues since, by definition, this is still a "single observation". This implies a fundamental limitation on the amount of information that can be recorded in our brain. In other words, it is possible that at $t_1$ we "observe" a lot more stuff than our brain retains. 

This limitation of the brain is due to the fact that a given configuration of particles $\vec{x}_2$ (which includes our brain) at $t_2$ can be reached from many different particle configurations $\vec{x}_1$ at $t_1$. The lack of one to one correspondence is the key to lack of perfect memory. Now, if we could access both $\vert \psi \rangle$ \emph{and} $\vec{x}$, then there would, in fact, be one to one correspondence since the dynamics of $(\vert \psi \rangle, \vec{x})$ is deterministic. There is no fundamental "quantum mechanical" reason for us not to be able to do that. We have simply set up a "classical" rule that we can't, since that is the only source of (classically) "probabilistic" nature of $\sigma$. 

Now, if we define our brain through $(\vert \psi \rangle, \vec{x})$, then it would, indeed have unlimited capacity. But, in this case, we are "allowed" only an access to a "part of" our brain, and thus only part of its memory. In this new language the picture is as follows. "We" ourselves do not consist of anything physical and, therefore, we have zero memory. Our brain, consisting of \emph{both} particles ($\vec{x}$) \emph{and} waves ($\vert \psi \rangle$) has absolutely perfect memory. However, we only have access to \emph{part} of our brain (for example, that part consists of \emph{complete} information about $\vec{x}$ and \emph{no} information about $\vert\psi \rangle$). Thus, by making repeated observations of our brain we are lead to believe that we have "imperfect memory". 

Let us now go back to the question we asked earlier: if we have a dynamic picture how $\rho$ can be truly static? Let us imagine the following scenario. We first perform an observation at $t=t_1$. We then predict the outcome for the time $t=t_2$. We then "record" both observation and the predicted outcome in our brain. At time $t=t_1 + \delta t$ we completely forget both the observation and predicted outcome. Then at a time $t=t_2$ we make another observation, in order to make a prediction for $t=t_3$. 

Now, at a time $t_2 - \delta t$ we have not made the observation \emph{yet}. This means that, at a time $t_2 - \delta t$ we were not looking at \emph{any} part of the universe, including our brain. This means that we have "forgotten" our observation at $t_1$. At the same time, however, we \emph{can} "remember" the "static equilibrium" $\rho (\vert \psi \rangle, \vec{x})$ without "looking" at our brain. Thus, at time $t_2 - \delta t$ we will write down $\rho$; then, at time $t_2$ we will "look" at everything (including our brain) and record $\vec{x} (t_2)$. And then we will use both $\vec{x} (t_2)$ as well as $\rho$ in order to make predictions for $t_3$. Since at a time $t_2 - \delta t$ we had no memory of the observation at $t_1$, we have "written down" the same $\rho$ as we did at $t_1 - \delta t$. That is the ultimate reason why $\rho$ is time independent. 

Now, since our observation at $t_2$ includes the observation of our brain that has a record of what happened at $t_1$, the above argument can be stated as follows. The probability distribution $\rho$ is an "a priori knowledge" that is "innate" to us independently of our memories (namely, the "knowledge of physics"). Then when we access our brain at $t_2$, we use our "a priori knowledge" in order to evaluate what we see (in particular, our "physics knowledge" would tell us how likely a wave would look a certain way for a given distribution of particles that we "see"). Thus, while the information we access is time dependent, the "physics knowledge" we have is not. The "static" probability distribution $\rho$ is precisely that knowledge. 

Now, if we could access both $\vec{x}$ \emph{and} $\vert \psi \rangle$, we would no longer need $\rho$. The only "physics knowledge" we would need is a \emph{deterministic} dynamics that would predict the future $(\vert \psi \rangle, \vec{x})$ with absolute certainty. The reason we need $\rho$ is precisely because we are not allowed to access $\vert \psi \rangle$, and we can \emph{only} access $\vec{x}$. Thus, $\rho$ is our "guide" on how to "infer" some information about $\vert \psi \rangle$ from $\vec{x}$ (even though that information is probabilistic). In terms of our brain, $\rho$ serves as a guide for us to "guess" the "rest of our brain" (described by $\vert \psi \rangle$) based on our reading of its part described by $\vec{x}$ . This means that we \emph{need} $\rho$ in order to access our memory. Thus, in order to avoid circular arguments, $\rho$ better be memory-independent, which is why it is time-independent. 

\subsection*{5. The "dynamic picture" predicted from "static equilibrium"}

In order to satisfy ourselves regarding the "static picture", let us "switch off" collapse mechanism (or, in other words, $\vec{x}$-dependence of $\vert \psi \rangle$) for a while and attempt to reproduce the standard "dynamic" Bohmian argument based on our "static" one. First, let us expand the divergence equation for "static equilibrium" in the following way:
\beq - \frac{\partial \rho}{\partial t} \Big\vert_{(\vert \psi \rangle, \vec{x})} = \vec{\nabla}_x (\rho \vec{v}_x) + \vert v_{\psi} \rangle \cdot \vec{\nabla}_{\psi} \rho + \rho \vec{\nabla}_{\psi} \cdot \vert v_{\psi} \rangle \eeq
Let us now imagine that there are many different particles in $(\vert \psi \rangle, \vec{x})$ space moving according to the same "deterministic law", but they have different initial conditions, and they do not interact with each other. Suppose we are "sitting" on one of these particles, and measuring the "density" of the other particles in its neighborhood. In this case, the latter will evolve as 
\beq \frac{\partial \rho}{\partial t} \Big\vert_{(\vert \psi (t) \rangle, \vec{x} (t))}= \frac{\partial \rho}{\partial t} \Big\vert_{(\vert \psi \rangle, \vec{x})} + \vert v_{\psi} \rangle \cdot \vec{\nabla}_{\psi} \rho \eeq
which means that the "static" equation we have written is equivalent to 
\beq \frac{\partial \rho}{\partial t} \Big\vert_{(\vert \psi (t) \rangle, \vec{x}(t))} = \frac{\partial \rho}{\partial t} \Big\vert_{(\vert \psi \rangle, \vec{x})} - \vec{\nabla}_x \cdot (\rho \vec{v}_x) - \rho \vec{\nabla}_{\psi} \cdot \vert v_{\psi} \rangle \eeq
Furthermore, we recall that "complex divergence" is given by 
\beq div \; F \; = \; 2 \; Re \; \Big( \frac{\partial F}{\partial z} \Big) \eeq
If we generalize it to multidimensional case (and further extrapolate it to uncountable dimensionality of $\vert \psi \rangle$ and $\vert v_{\psi} \rangle$) then we will find that the presence of a coefficient $i$ in the unitary evolution equation, 
\beq \vert v \rangle = -iH \vert \psi \rangle \eeq
implies that
\beq \vec{\nabla} \cdot \vert v_{\psi} \rangle = 0 \eeq
as long as $H$ is Hermitian. This means that we can "throw away" the last term from the right hand side of the equation for $d \rho /dt$. Thus, we obtain  
\beq \frac{\partial \rho}{\partial t} \Big\vert_{(\vert \psi (t) \rangle, \vec{x}(t))} = \frac{\partial \rho}{\partial t} \Big\vert_{(\vert \psi \rangle, \vec{x})} - \vec{\nabla}_x \cdot (\rho \vec{v}_x) \eeq
Now, if we substitute a \emph{static equilibrium equation}, 
\beq \frac{\partial \rho}{\partial t} \Big\vert_{(\vert \psi \rangle, \vec{x})} = 0 \eeq
which we are "advertising", we then obtain 
\beq \frac{\partial \rho}{\partial t} \Big\vert_{(\vert \psi (t) \rangle, \vec{x}(t))} =  - \vec{\nabla}_x \cdot (\rho \vec{v}_x), \eeq
which exactly coincides with \emph{dynamic equilibrium} equation we are used to in the context of random walk in $\vec{x}$ \emph{alone}. From this point on the argument is a carbon copy of the standard Bohmian one: if we cleverly define $\vec{v}_x$ in such a way that 
\beq \frac{\partial \vert \psi \vert^2}{\partial t} = - \vec{\nabla}_x \cdot (\vert \psi \vert^2 \vec{v}_x) \eeq
is satisfied, then $\rho = \vert \psi \vert^2$ is an "equilibrium point" of the theory. 

We notice, however, that the above argument crucially depends on the presence of $i$ in the evolution equation. In particular, we have appealed to the presence of $i$ when we "got rid" of unwanted $\vec{\nabla} \cdot \vec{v}$ term, by identifying it with zero. There are two issues that arise from this. First of all, if we introduce a collapse mechanism in a form of multiplication by Gaussian, this would \emph{not} have the "$i$" that the ordinary unitary evolution equation does. Secondly, and perhaps even more importantly, the "logic" behind \emph{either} dynamic \emph{or} static equilibrium argument is independent of our knowledge of "physics", which includes our knowledge of unitary evolution equation. Thus, \emph{if} no "logical mistakes" were made in either argument, their results should match, even in the case of "wrong physics" that lacks unitarity. The fact that such is not the case seems to suggest that one of these arguments is "wrong". If so, we have to find out which one. 

One source of mismatch between the two probabilities is that in a static case we are considering \emph{overall probability}, while in the "dynamic case" we are considering a \emph{conditional probability}, $\rho / \sigma$ where 
\beq \sigma (\vert \psi \rangle) = \int d^{3N} x \; \rho (\vert \psi \rangle, \vec{x}) \eeq
Let us, therefore, rewrite our "static equation" for the conditional probabilities, as well. We recall that, right \emph{before} we used "zero divergence" our equation for $\rho$ took the form 
\beq \frac{\partial \rho}{\partial t} = - \vec{\nabla}_x \cdot (\rho \vec{v}_x) - \rho \vec{\nabla}_{\psi} \cdot \vert v_{\psi} \rangle, \eeq
Here, we have dropped one of the time derivative terms, because we are assuming from now on that static equilibrium has already been reached. Now, if we "integrate" it over $\vec{x}$, we obtain
\beq \frac{\partial \sigma}{\partial t} = \int d^{3N} x \; (- \vec{\nabla}_x \cdot (\rho \vec{v}_x) - \rho \vec{\nabla}_{\psi} \cdot \vert v_{\psi} \rangle ) \eeq
Now, by Gauss' theorem, the first term on the right hand side is equal to the surface integral of $\rho \vec{v}_x$ over a "very large" sphere that encircles "all of the $\vec{x}$-space". Thus, if we assume a "boundary conditions" that $\vec{v}_x =0$ at $\vec{x} = \infty$, then the first term drops out, and the equation becomes 
\beq \frac{\partial \sigma}{\partial t} = - \int d^{3N} x \; \rho \vec{\nabla}_{\psi} \cdot \vert v_{\psi} \rangle  \eeq
Let us now write the expression for the time derivative of "conditional probability" $\rho / \sigma$:
\beq \frac{\partial}{\partial t} \frac{\rho}{\sigma} = \frac{1}{\sigma} \frac{\partial \rho}{\partial t} - \frac{\rho}{\sigma^2} \frac{\partial \sigma}{\partial t} \eeq
Now, while one of the two terms of $\partial \sigma / \partial t$ was killed out by the integral, that term is not zero \emph{outside of integration}. Thus, the expression for $\partial \rho / \partial t$ has both terms. If we now substitute "two terms" in place of $\partial \rho / \partial t$, and "one term" in place of $\partial \sigma / \partial t$, we obtain
\beq \frac{\partial}{\partial t} \frac{\rho}{\sigma} = \frac{1}{\sigma} (- \vec{\nabla}_x \cdot (\rho \vec{v}_x) - \rho \vec{\nabla}_{\psi} \cdot \vert v_{\psi} \rangle) - \frac{\rho}{\sigma^2} \Big(- \int d^{3N} x \; \rho \vec{\nabla}_{\psi} \cdot \vert v_{\psi} \rangle   \Big) \eeq
In light of the fact that $\sigma$ is defined as an integral over $\vec{x}$, it is a function of $\psi$ alone and, therefore, we are free to move $\sigma$ in and out of differentiation and integration over $\vec{x}$. Thus, we can rewrite the above equation as follows
\beq \frac{\partial}{\partial t} \frac{\rho}{\sigma} = - \vec{\nabla}_x \cdot \Big(\frac{\rho}{\sigma} \vec{v}_x \Big) - \frac{\rho}{\sigma} \vec{\nabla}_{\psi} \cdot \vert v_{\psi} \rangle + \frac{\rho}{\sigma} \Big( \int d^{3N} x \; \frac{\rho}{\sigma} \vec{\nabla}_{\psi} \cdot \vert v_{\psi} \rangle   \Big) \eeq
which can be further rewritten as 
\beq \frac{\partial}{\partial t} \frac{\rho}{\sigma} = - \vec{\nabla}_x \cdot \Big(\frac{\rho}{\sigma} \vec{v}_x \Big) - \frac{\rho}{\sigma} \Big( \vec{\nabla}_{\psi} \cdot \vert v_{\psi} \rangle - \int d^{3N} x \; \frac{\rho}{\sigma} \vec{\nabla}_{\psi} \cdot \vert v_{\psi} \rangle   \Big) \eeq
The last two terms on the right hand side represent the difference between the value of $\vec{\nabla}_{\psi} \vert v_{\psi} \rangle$ for \emph{a given $\vec{x}$} and its "average" over all $\vec{x}$ (but with fixed $\psi$). In order to "set" this difference to zero we have to make sure that $\vert v_{\psi} \rangle$ is $\vec{x}$-independent. Remembering that $\vec{x}$ represents position of a "particle", while $\vert v_{\psi}$ represents velocity of a "wave", this is equivalent to saying that the behavior of a wave is independent of the way the particle influences it. 

Now, we have been saying earlier that \emph{any} non-unitary influence on the wave would cause a problem, not just the one coming from the particle. The way particle-independent non-unitary influences are "dealt with" is through normalization factor in the definition of "conditional probability". If non-unitary influence is $\vec{x}$-independent, its impact on pointwise $\rho$ will be identical to its impact on normalization factor (since the latter is an "average" of a former); thus, upon "dividing" the former by the latter we cancel the non-unitary effects. Particle-dependence (or in other words $\vec{x}$-dependence) is precisely what prevents us from doing such cancellation. 

This immediately tells us what is the root of "mismatch" between statical and dynamical predictions. Namely, it is the very same thing we were talking about back in Chapter 2. Just to repeat the Chapter 2 argument, we have noticed that dynamical picture assumes that we "know" $\psi$ and, at the same time, "not know" $\vec{x}$. This implicitly assumes $\vec{x}$-independence of $\psi$. \emph{Only} when such assumption is true do we expect the equation produced by "dynamical picture" to hold. In the latter case the "dynamically-produced" equation, in fact, agrees with 'statically-produced" one. In all other cases, the dynamical picture no longer works; but we now have a static picture to "fill in a gap". 

However, in light of the fact that divergence "sums" different components, it is possible to hope that $\vec{\nabla}_{\psi} \cdot \vert v_{\psi} \rangle$ is $\vec{x}$-independent, despite $\vec{x}$-dependence of different "terms" in the "sum". Let us, therefore, attempt to evaluate this divergence to see if that is, in fact, the case. Clearly, $\vec{\nabla}_{\psi} \vert v_{\psi} \rangle$ on the right hand side involves uncountably many degrees of freedom. We can identify their "sum" with corresponding "integral" multiplied by "infinitely large" factor $K$: 
\beq \vec{\nabla}_{\psi} \cdot \vert v_{\psi} \rangle = K \int d^{3N} y \; \frac{\delta}{\delta \langle y \vert \psi \rangle} \langle y \vert v_{\psi} (\psi, x) \rangle, \eeq
where $x$ is a position of a "beable particle" while $y$ is a "dummy index" representing "arbitrary component" of $\psi$ as we "take divergence". Thus, $\langle y \vert v_{\psi} (\psi, x) \rangle$ represents an effect on the value of $\psi$ around $\vec{y}$ as a result of "collapse" produced by the particle at $\vec{x}$. The definition of $\vert v_{\psi} (\psi, x) \rangle$ is a "blank space" that is to be "filled in" by a "collapse model". Let us now assume that our "collapse model" produces $\vert v_{\psi} \rangle$ in a form
\beq \vert v_{\psi} (\psi, x) \rangle = a \; \int d^{3N} y \; f(\vec{y} - \vec{x} ) \vert y \rangle \langle y \vert \psi \rangle \eeq
where $a$ is some coefficient. If we substitute this to the right hand side of precious equation, we obtain
\beq \vec{\nabla}_{\psi} \cdot \vert v_{\psi} \rangle = aL \int d^{3N} y \; f (\vec{y} - \vec{x}) \eeq
where $L = K \langle y \vert y \rangle =K \delta^{(3N)} (0)$ is another "infinite" constant. The expression on the right hand side is, indeed, $\vec{x}$-independent. Thus, if we substitute this into 
\beq \frac{\partial}{\partial t} \frac{\rho}{\sigma} = - \vec{\nabla}_x \cdot \Big(\frac{\rho}{\sigma} \vec{v}_x \Big) - \frac{\rho}{\sigma} \Big( \vec{\nabla}_{\psi} \cdot \vert v_{\psi} \rangle - \int d^{3N} x \; \frac{\rho}{\sigma} \vec{\nabla}_{\psi} \cdot \vert v_{\psi} \rangle   \Big), \eeq
then the last two terms cancel, resulting in the expression
\beq \frac{\partial}{\partial t} \frac{\rho}{\sigma} = - \vec{\nabla}_x \cdot \Big(\frac{\rho}{\sigma} \vec{v}_x \Big). \eeq
which matches the one we obtained in the scenario free of particle's influence. 

\subsection*{6. The additional influence on $\psi$: unitary and non-unitary alternatives}

While at the end of the previous section we seem to have arrived with the "desired" equation for $\rho$, there is a little catch there: $\vert \psi \vert^2$ no longer obeys that equation, since $\psi$ has an additional influence from the position $\vec{x}$ of the particle. In order to conclude that $\rho = \vert \psi \vert^2$ is an equilibrium point we need \emph{both} sides to obey the same dynamics. Yet, up till now we just took it for granted that $\psi$ would be "taken care of" and we have focused exclusively on $\rho$.

Now, there are two alternative ways of making sure that $\vert \psi \vert^2$ continues to obey the dynamics we want it to obey. One such way is to simply "do this by hand". In other words, we no longer appeal to the probability current or any other physical intuition in deriving Pilot Wave model. Instead, we simply use "inverse Laplassian" for the "probability current".  We recall that our proposal of "continuous collapse" was 
\beq \frac{\partial \psi}{\partial t} = -i H \psi - \frac{a}{2} (\vec{x} - \vec{x}_B)^2 \psi \eeq
which implies that 
\beq \frac{\partial \vert \psi \vert^2}{\partial t} = - a \vert \vec{x} - \vec{x}_B \vert^2 \vert \psi \vert^2 \eeq
which implies that the above can be satisfied if we set the probability current to be
\beq \vec{j} = -a \nabla^{-2} (\vert \vec{x} - \vec{x}_B \vert^2 \vert \psi \vert^2 ) \eeq
which, in turn, can be satisfied by setting Pilot Wave model to be 
\beq \vec{v}_x = - \frac{a}{\vert \psi \vert^2} \vert \vec{x} - \vec{x}_B \vert^2 \vert \psi \vert^2  \eeq
This approach was, in fact, used both in \cite{Epstein} and in \cite{Sverdlov} for other purposes. For example, in \cite{Sverdlov} the "non-locality" was introduced for the purposes of incorporating creation and annihilation of particles which, due to its discreteness, is "non-local" in nature and, therefore, is difficult to accommodate through "gradient". 

However, at least in non-relativistic case we have no other reason of introducing such non-locality. So it makes sense to ask ourselves whether or not we can avoid it. In fact, we can! The dynamics proposed above can be rewritten as 
\beq V' = V - ia (\vec{x} - \vec{x}_B)^2 \eeq
Now, we already know that the Pilot Wave model, as defined by Bohm, works for the case of \emph{real} potential:
\beq V \in \mathbb{R} \; \Longrightarrow \; \vec{v} = \frac{1}{m} \; \vec{\nabla} \; Im \; ln \; \psi \; {\rm WORKS} \eeq
 The "complex value" is a problem since the evolution is no longer unitary and, therefore, can not be interpreted as "probability current flow" which is what Bohmian mechanics is bases upon. The way to "fix" it is simply to replace $-ia$ with $+a$: 
\beq V' = V + a (\vec{x} - \vec{x}_B)^2 \eeq
on physical grounds we would expect the "high potential" to "keep the wave away" from the "far away" region from the particle. At the same time, if $a$ is very small its effect is negligeable as long as the distance is "reasonable"; but, at the same time, it would become important in the event of entanglement since the distance in \emph{configuration space} would become "large". This, in fact, is the only sought-after outcome of GRW collapse scenario and, therefore, can replace the latter in our theory. 

Incidentally, setting $V$ to be real allows us to avoid "uncountable divergence" associated with $\vec{\nabla}_{\psi} \cdot \vert v_{\psi} \rangle$. If we recall that complex divergence is given by 
\beq div \; F \; = \; 2 \; Re \; \frac{\partial F}{\partial z} \eeq
we will see that by setting $H$ to be Hermitian (which is generalization of "real"), $iH$ will be a generalization of "imaginary", which would set the above real parts (which form "uncountably many terms") to zero. This means that $\vec{\nabla}_{\psi} \cdot (\rho \vert v_{\psi} \rangle)$ will reduce to $\vert v_{\psi} \rangle \cdot \vec{\nabla} \rho$ which can be thought of as "one dimensional" derivative, as opposed to "infinite dimensional" one. Thus, we would be able to get rid of uncountable divergences in a lot less sloppy way, without having to resort to "uncountably large coefficients" that we have been using. 

There is, however, a question that one can ask: if we change the factor of $i$, then our dynamic equation becomes 
\beq \frac{\partial \psi}{\partial t} = -i H \psi + \frac{ia}{2} (\vec{x} - \vec{x}_B)^2 \psi \eeq
which seems to be saying that additional term is only a phase factor. The good news, however, is that the same question can be asked within the bounds of any textbook quantum mechanics problem: why would the probability of finding a particle within a potential well be "very small" if the potential only contributes to the unitary phase factor? We can, therefore, satisfy ourselves by simply answering that quantum mechanics question within the more familiar context we have just referred to. 

The above question, of course, can be easily answered within the context of time independent Schrodinger's equation. In time dependent context it is a bit more tricky for the above stated reason. However, if we apply$-iH \delta t$ \emph{twice} we will get some clues. After we apply $-iH\delta t$ the \emph{first time}, the magnitude $\vert \psi \vert$ would stay unaffected, \emph{but} the $\partial^2 \psi / \partial x^2$ would pick up "imaginary component" relative to $\psi$. Since the second derivative is part of Hamiltonian, we would conclude that $H$ would now have imaginary part as well. Thus, $iH$ would now have a "real part".Thus, after applying $-iH \delta t$ the \emph{second time} we would, in fact, affect magnitude. 

Now, consider one dimensional scenario. If $\partial^2 V / \partial x^2>0$, and if we assume that we have "started out" from $\psi$ being purely real and positive, this would imply that the sign of $\partial^3 \psi /\partial t \partial x^2$ is $-i$. Thus, after "very small" time interval, $\partial^2 \psi / \partial x^2$ will pick up a small $-i$ component.Thus, kinetic part of Hamiltonian would pick a small $+i$ component. Thus,$\partial \psi / \partial t$, being proportional to $-iH$ would pick a small $+1$ component. Now, the "potential barrier" is characterized by$\partial^2 V / \partial x^2 <0$ which means that $\partial \psi /\partial t$ will pick $-1$ sign, as desired. 

The "time reversal" symmetry is taken care in the following way. We have assumed that "at the beginning" $(t=t_0)$ there is no phase factor. This,however, implies that at $t<t_0$ there is a phase factor "in the opposite direction". Thus, as we go "back in time" we will find our wave function decreasing at the places of high potential. Equivalently, if we go"forward in time" then, before $t=t_0$, the wave function has increased to the "original value" and then later started to decrease. Thus, the "time reversal" symmetry is broken down simply because we have assumed that we are living in $t > t_0$. Again, we have concluded that at $t> t_0$ the time derivative has the same sign as the second space derivative (which implies increase at the low potential and decrease at the high potential),while at $t<t_0$ the situation is the opposite. 

This, however, raises another question: what if potential increases away from the center to infinity, while having positive second derivative all along? Our previous argument seems to imply that the wave function would increase away from the center while physically we would expect the opposite to be the case. This question can be answered by noticing that in our qualitative analysis we were exclusively focused on phase and,therefore, assumed that amplitude is constant in space at $t=t_0$. If such is not the case, then \emph{despite} the $-i$ factor in $\partial^2 (e^{i\theta})/ \partial x^2$, if we include amplitude ($r$) we might well have$+i$ in $\partial^2 (r e^{i \theta})/ \partial x^2$. 

It is easy to see that it is the logarithm of the coefficient that plays the main role here and. Thus, the fact that logarithm rapidly becomes"very large" with the negative sign towards the boundary of the support of the wave function implies that the coefficient there will, in fact, be $+i$, which means that the function near the boundary will further decrease instead of increasing and we have avoided our problem. This argument, however, relies on the fact that the support of the wave function is, in fact finite. 

While the assumption of finite support seems to be justified by normalization, it doesn't have to be the case, as exemplified by "fixed momentum" particle. In the cases of unbounded support we would, in fact,get exotic cases of further increase of wave function we have discussed earlier. Studying these cases is beyond the scope of this paper; but we will investigate these in the upcoming paper (see \cite{oscillator}). for our present purposes, we can be content with "cosmological" assumption that the support of wave function happened to be bound "at the beginning of the universe". 

It is interesting to note that our analysis could be copied for a "more standard" GRW theory that involves discrete hits. In this case, we would conclude that a "hit" does not have to be a multiplication by $e^{-ax^2}$as typically assumed. Instead, it could be a multiplication by$e^{-iax^2}$. In this case, the "hit" itself would have no influence on the magnitude of a wave \emph{right after} its occurrence; \emph{but} it would set its "phase factor" in such a way that \emph{subsequent  Schr"/odinger's evolution} would result in the desired decrease of the amplitude, similar to the one we have discussed. 

The reason in standard GRW approach this was not done is probably because discrete hits don't have much in common with Hamiltonian \emph{even if}the extra $i$-factor was inserted; thus, it is logical to "make things simple" by using $e^{-ax^2}$ since there is not much to be gained by being"clever". In our case, however, by replacing discrete hits with a continuous process, we have made our picture "almost the same" as Hamiltonian; the only difference is precisely that $i$ factor. This factor, further, place deciding role between us using "ugly" $\nabla^{-2}$ in Pilot Wave model versus "beautiful" $\vec{\nabla}$ originally proposed by Bohm. For these reasons we choose to introduce this factor, unless we are dealing with models where we need $\vec{\nabla}^{-2}$ regardless (such as \cite{Epstein} and \cite{Sverdlov}). 

\subsection*{7. Momentum conservation}

As was stated in the before, the definition of $\vert v_{\psi} \rangle$ amounts to specifying a "collapse mechanism". We have further found out that our picture would provide the desired result as long as collapse mechanism is "translationally invariant"; or, in other words, if the particle is located at $\vec{x}$, then its influence on the behavior of the wave around $\vec{y}$ is a function strictly of $\vec{y} - \vec{x}$. As long as the latter is true, we do get the desired result, regardless of the specifics of the mechanism.

In light of this, we can, in principle, say that we are done and leave the specification of collapse mechanism to the reader. However, I believe that it is worth elaborating on the mechanism for reasons independent of the main subject of the paper. In particular, if we use the mechanism presented by GRW models, and apply it to a single harmonic, then its multiplication by Gaussian would produce harmonics  of other frequencies; thus, "conservation of momentum" would be violated. 

In principle, this does not disprove a model: after all, we have never "seen" momentum. We have only seen an arrow of measuring apparatus pointing in a certain direction (which signifies a certain momentum). That direction, of course, is defined in "position space". Thus, one might argue that as long as GRW model makes correct predictions in "position space", it can predict all of the "momentum measurements" and find them consistent with predictions of quantum mechanics. Since stadard quantum mechanics predicts that momentum conservation "holds in the lab", so does GRW model, despite the subtle contradictions with momentum conservation in the setup. However, I still believe that the models that respect momentum conservation are a lot better aesthetically since only in the latter case we can explicitly include the "momentum exchange between particle and a measuring device" into the model at hand. 

Consider the following scenario. Suppose we first measure momentum of a particle. Then we measure its position, and then, again measure momentum the second time. The second momentum observation, of course, does not match the first one. This can be explained through the "momentum exchange" the particle has with measuring device. This "momentum exchange" is the result of the fact that, in order to measure position, we have to "localize" measuring device in space. Thus, by uncertainty principle, the latter has uncertain momentum. That is a reason why it can pass "uncertain momentum" onto a particle. On the other hand, if we had "less precise" position measurement, we would not have to "localize" measuring device as much. Thus, the momentum of the latter would be less uncertain, which is why the momentum it "passes" onto a particle would be less uncertain as well. 

Now, in the above paragraph we were appealing to the "exchange of momentum" between the particle and measuring apparatus. But, if momentum could be created through multiplication by Gaussian, then the discussion about "momentum exchange" becomes relatively meaningless, since the notion of "exchange" intrinsically appeals to "conservation". Thus, in order to "save" our "momentum exchange" argument, we have to redefine $V(\vec{y}- \vec{x})$ in such a way that it is no longer a "source of momentum" but, rather, a "catalyst of momentum exchange". Here, it is understood that by $V$ we typically mean either "real" or "imaginary" options discussed in previous section:
\beq V = -\frac{ax^2}{2} \; {\rm OR} \; V = -i\frac{ax^2}{2} \eeq
but, since our argument only depends on $V$ being "very large" at the "large distances" we will leave it in a general form. 

We recall from Noether's theorem that momentum conservation is equivalent of translational covariance. The reason $V (\vec{y} - \vec{x})$ violates the former is that it violates translational covariance by setting $\vec{x}$ as "preferred value of $\vec{y}$". Strictly speaking, within the "particle-wave" context, the translational symmetry was \emph{not} violated since $\vec{x}$ is simply a position of a particle which means that it is not "better off" than any other location would have been if the particle were to move there. However, the context of Noether's theorem is wave-alone as opposed to particle+wave. In wave-alone context we \emph{do} violate the translational symmetry. Furthermore, we are \emph{not} in a position to "redo" Noether's theorem for particle+wave context, since the "momentum conservation" would require point-like excitations of the wave to "balance" the point-like momentum changes produced by particle's acceleration.  

What we can do instead is to take advantage of the fact that $\vec{x}$ represents a \emph{configuration of particles} (as opposed to one single particle) and, therefore, replace the dependence on $\vec{x} = (\vec{x}_1, \cdots \vec{x}_N)$ with a dependence on a \emph{set of vectors},  
\beq \{ \vec{x}_j - \vec{x}_i \vert 1 \leq i, j \leq N \}, \eeq
where $N$ is the total number of particles a "configuration space" is meant to describe. The "change" of $\vec{x}_i - \vec{x}_j$ refers to "simultaneous translation" of $\vec{x}_i$ and $\vec{x}_j$ in such a way that their center of mass stays fixed. It is easy to see that fixed center of mass implies the fixed total momentum, while variation of $\vec{x}_i - \vec{x}_j$ implies momentum exchange. Furthermore, in order to avoid the violation of angular momentum conservation as well, we have to demand the rotational symmetry. This can be done by simply replacing $\vec{x}_i - \vec{x}_j$ with $\vert \vec{x}_i - \vec{x}_j \vert$. Thus, we propose the definition of $f$ in the following form:
\beq V (\vec{y} - \vec{x}) = \sum_{i, j} U(\vert \vec{x}_j - \vec{x}_j \vert, \vert \vec{y}_j - \vec{y}_j \vert), \eeq
where $g (\rho, \sigma)$ is designed in such a way that it has a "minimum" when $\rho = \sigma$ and stays "very close" to the minimum whenever $\vert \rho - \sigma \vert$ is "neither small nor large"; but then $g(\rho, \sigma)$ "explodes" when $\rho - \sigma$ becomes "large". This would allow us to "mimic" GRW argument: the effects of $g$ are negligible unless they are "magnified" through entanglement. At the same time, the momentum conservation will be respected. 

In the above equation we refrained from replacing $U (\rho, \sigma)$ with $U (\sigma - \rho)$ for the following reason. We would like the momentum exchange to be "stronger" between the particles "nearby" than between the ones "far away". Now, the former is represented by both $\rho$ and $\sigma$ being "small", while the latter is represented by them both being "large". Thus, if $\rho$ and $\sigma$ are relatively small, while $\Lambda$ is "very large", we would expect that 
\beq U( \rho + \Lambda, \sigma + \Lambda) \ll U (\rho, \sigma) \eeq
Unfortunately, however, we are unable to remove non-locality completely. While we could set $U (\rho, \sigma) =0$ for "very large" $\rho$ and $\sigma$, we have to allow it to be non-zero when the latter are "small enough"; thus, our momentum exchange is at best "quasi-local". This situation, however, is not any worse from the non-locality we need to adopt \emph{anyway} in the definition of configuration space in order to write down either Schr"/odinger's equation or Pilot Wave model, to begin with. 

\subsection*{8. Conclusion}

In this paper we have combined Bohmian particles and GRW collapse model into a single theory, where Bohmian particle serves as a "center" around which a continuous GRW-type collapse occurs. This seems to improve things on both ends. On GRW end, we no longer need to impose a "spontaneous collapse" (as was done in \cite{GRW1} and \cite{GRW2}), nor do we need to come up with any additional mechanism of deterministic continuous collapse (\cite{continuous}). On the Bohmian end, we no longer need to rely on "decoherence" argument which was disputed by Leggett (\cite{Leggett1} and \cite{Leggett2}) and which, if true, can result in "traveling between universes" in non-local Pilot Wave models (\cite{Epstein}, Chapter 3, and \cite{Sverdlov}). 

This work is a direct result of my private discussions with Ward Struyve regarding the "collapse mechanism" I proposed in \cite{Sverdlov}. He has suggested to replace that mechanism with "spontaneous collapses" of GRW model, centered around Bohmian particle. However, I found that idea to violate Ocams razor since Bohmian particle doesn't seem to play any role in "improving upon" GRW model, which makes its presence in the theory unnatural. However, I decided that this conflict with Ocam's razor can be fixed if Bohmian particle is a source of \emph{continuous} localization, which would furnish it with far more significant role in the theory. Incidentally, this would also harmonize with my deterministic world view a lot more than spontaneous collapse models (with or without Bohmian particle). 

In future work it might be interesting to compare some finer implications of a choice of using Bohmian particle as a "continuous collapse" source as contrasted with other \emph{deterministic} continuous collapse sources (such as presented in \cite{continuous}). For example, the particle follows a continuous one-dimensional trajectory, which can not be said about other kinds of deterministic sources of continuous collapse. This might result in a different kinds of "resonances" produced through the continued collapse process which might either strengthen or weaken different aspects of collapse.

Another offshoot for future work is to explore the modifications of GRW mechanism proposed in Chapters 6 and 7 independently of a "bigger context" of this paper. In particular, \emph{despite} the fact that I favor "continuous collapse" scenario, it might be still interesting to apply Chapters 6 and 7 to the "stochastic" localization events, as originally introduced in \cite{GRW1} and \cite{GRW2}. In particular, it might be interesting to add extra $i$ to the "stochastic localization" hit (which would parallel Chapter 6) and also re-define "stochastic localization" hit in such a way that it would conserve momentum (which would parallel Chapter 7). 

I believe both of these are logical things to do: neither extra $i$ nor momentum conservation are "logically related" to a belief in continuous collapse; therefore, both should be adaptable to a discrete collapse settings. I believe it is important to separate logically unrelated modifications to theories and explore the effects of each such modification on its own which would enable us not to confuse the "ups" and "downs" of each such modification with the "ups" and "downs" that would arise from other ones. After that, of course, it might be logical to try different combinations of "including" some modifications and not others. This will be one of the things I will investigate in future work. 

{\bf Acknowledgement:} I would like to thank Ward Struyve for suggestion that Bohmian particle is a source of stochastic collapse which ultimately lead me to think of a "continuous and deterministic" version of his suggestion in the form of a present paper.

\end{document}